\documentclass[
 aps, pra,
 amsmath,amssymb,
 11pt,
 final,
tightenlines,
 twoside,
 onecolumn,
 nofloats,
nofootinbib,
 superscriptaddress,
 ]
{revtex4}

\usepackage[T2A]{fontenc}
\usepackage[utf8x]{inputenc}
\usepackage[russian,english]{babel}
\usepackage{graphicx}
\usepackage{dcolumn}
\usepackage{bm}

\usepackage{ulem}

\input{maik.rty}

\setcitestyle{authoryear,round}
\def\squareforqed{\hbox{\rlap{$\sqcap$}$\sqcup$}}

\def\sq{\ifmmode\squareforqed\else{\unskip\nobreak\hfil
\penalty50\hskip1em\null\nobreak\hfil\squareforqed
\parfillskip=0pt\finalhyphendemerits=0\endgraf}\fi}

\def\utw{\smash{\rlap{\lower5pt\hbox{$\sim$}}}}

\def\udtw{\smash{\rlap{\lower6pt\hbox{$\approx$}}}}

\def\diameter{{\ifmmode\mathchoice
{\ooalign{\hfil\hbox{$\displaystyle/$}\hfil\crcr
{\hbox{$\displaystyle\mathchar"20D$}}}}
{\ooalign{\hfil\hbox{$\textstyle/$}\hfil\crcr
{\hbox{$\textstyle\mathchar"20D$}}}}
{\ooalign{\hfil\hbox{$\scriptstyle/$}\hfil\crcr
{\hbox{$\scriptstyle\mathchar"20D$}}}}
{\ooalign{\hfil\hbox{$\scriptscriptstyle/$}\hfil\crcr
{\hbox{$\scriptscriptstyle\mathchar"20D$}}}}
\else{\ooalign{\hfil/\hfil\crcr\mathhexbox20D}}%
\fi}}

\def\be{\begin{equation}}
\def\ee{\end{equation}}
\def\ba{\begin{eqnarray}}
\def\ea{\end{eqnarray}}

\def\msun{M_\odot}
\def\lsun{L_\odot}

\def\ltsima{$\; \buildrel < \over \sim \;$}
\def\simlt{\lower.5ex\hbox{\ltsima}}
\def\gtsima{$\; \buildrel > \over \sim \;$}
\def\simgt{\lower.5ex\hbox{\gtsima}}

\usepackage[dvipsnames]{color}

\begin{document}

\selectlanguage{english}


\title{Far-Infrared Emission from a Late Supernova Remnant in an Inhomogeneous Medium\footnote{published in Astrophysical Bulletin, v.~80, pp.~22--37 (2025), doi:  10.1134/S199034132460090X}}

\author{\firstname{S.~A.}~\surname{Drozdov}}
 \email{sai.drozdov@gmail.com}
 \affiliation{Lebedev Physical Institute, Russian Academy of Sciences, Moscow, 117997 Russia}

\author{\firstname{S.~Yu.}~\surname{Dedikov}}
 \email{s.dedikov@asc.rssi.ru}
 \affiliation{Lebedev Physical Institute, Russian Academy of Sciences, Moscow, 117997 Russia}

\author{\firstname{E.~O.}~\surname{Vasiliev}}
 \email{eugstar@mail.ru}
 \affiliation{Lebedev Physical Institute, Russian Academy of Sciences, Moscow, 117997 Russia}

\begin{abstract}
The interstellar dust grains are swept up during the expansion of the supernova (SN) remnant, they penetrate behind the shock front, where they are heated and destroyed in the hot gas. This leads to a change in emissivity of such grains. In this work, we consider the evolution of the infrared (IR) luminosity of the SN remnant expanding into an inhomogeneous interstellar medium with lognormal distribution of the density fluctuations. The IR luminosity of the swept-up interstellar dust rapidly increases during the first several thousand years after the SN explosion, and reaches the maximum value. Afterwards, it decreases due to the destruction of the dust grains in hot gas and their declining emissivity in the cooling down gas of the shell. We show how the IR luminosity of dust in the SN remnant depends on the dispersion of the gas density in front of the SN shock front. We find that for the significant period of time (40 -- 50 kyr) the maximum of the dust IR luminosity peaks at the range centered at 70$\mu$m. Therefore, this band can be considered as the most optimal range for studying the late SN remnants. We illustrate that during evolution, the dust temperature changes from 70 to 20~K, and only slightly depends on the inhomogeneity of the medium. In the radiative phase, the strong emission lines of metal ions emerge above the dust continuum. Their luminosity rapidly increases and exceeds the dust continuum luminosity by $\sim 10-10^3$ times. The point in time when the high luminosity in the lines is reached strongly depends on the inhomogeneity of the medium. We discuss possibilities for detection of the IR emission both in dust continuum and in lines. We expect that their ratios will allow to estimate the inhomogeneity of the medium, where the remnant is expanding.
\end{abstract}

\maketitle

\section{INTRODUCTION}

Interstellar dust is produced in significant quantities in ejecta of Type II supernovae \citep{Todini2001,Nozawa2003,Sarangi2015,Sluder2018}. Many observations of SN remnants several hundred years old reveal strong IR emission, which can definitely be associated with recently formed dust grains  \citep{Dwek1992,Williams2006,Matsuura2022,Priestley2022}. 

On the other hand, during the expansion of the SN remnant, interstellar dust is collected by its shell and is destroyed by strong shocks coming from the SN \citep{Jones1996,Bocchio2014,Micelotta2016}, in the processes of thermal and kinetic sputtering in hot gas with $T\simgt 10^6$К \citep{Barlow1978,Draine1979a,Draine1979b} and shattering during collisions of dust grains with each other in dense regions \citep{Borkowski1995,Jones1996,Bocchio2016}. Due to their inertia, dust particles can penetrate far behind the shock front and fall into hotter gas, spend several tens of thousands of years there, and effectively be destroyed during this period \citep{Slavin2020,vs2024}. Cooling by dust in hot gas can apparently significantly affect gas thermal evolution \citep{Ostriker1973,Smith1996}. A measure of this influence is the ratio of fluxes in the infrared and X-ray ranges \citep{Dwek1987a,Dwek1987b}. The observed ratio for SN remnants turns out to be significantly lower than the theoretical one \citep{Seok2015,Matsuura2022}, which can be explained by destruction of dust in the hot gas or by variation of the dust abundance in the gas ahead of the shock front \citep{Seok2015}.

After just a few thousand years of expansion of the SN remnant, the mass of the swept-up dust can reach several solar masses \citep{Slavin2020,vs2024,Dedikov2024}, which significantly exceeds the mass of dust produced in the SN, which is estimated to be less than $1\msun$ \citep{Matsuura2011,Matsuura2015,Wesson2021,Rho2008,Barlow2010,DeLooze2017,Niculescu-Duvaz2021,Gomez2012,Temim2013,DeLooze2019,Stanimirovic2005,Sandstrom2009}. Further the contribution from the swept-up interstellar dust in the total IR emission of the remnant apparently becomes more remarkable, if not dominant. It is obvious to assume that the produced dust should be hotter than the swept-up dust. However, the interstellar dust penetrating far behind the front also ends up in the hot gas, and its mass becomes greater than the injected one after several thousand years of the remnant’s evolution. Therefore, its contribution to the IR luminosity of the remnant will increase. The swept-up dust is located in a thick layer behind the shock front, the size of this layer gradually decreases due to decreasing expansion velocity of the shell and decelerating dust in it \citep{Slavin2020,vs2024}. Thus, dust grains gradually end up in a gas with a lower temperature and their effective sputtering ceases. The cooling of hot gas during expansion of the SN shell depends in general on energy of the explosion and properties of the environment, more precisely, on density and metallicity of the gas, and magnitude of its inhomogeneity.

Owing to the IR surveys of the Galaxy and the Magellanic Clouds \citep{Arendt1989,Saken1992,Ita2008,Kato2012,Pinheiro2011,Seok2013,Chawner2019,Chawner2020,Millard2021,Matsuura2022} our understanding of the IR emission properties, morphology, and evolution of SNe remnants has improved significantly. Most of the observed remnants are younger than a few thousand years {\citep[e.g.,][]{Milisavljevic2024}} and hence their IR emission is likely to be due to dust produced in the SN. However, several remnants are older, so the swept-up interstellar dust is expected to determine their IR luminosity. Its value should depend on the evolution of the hot gas mass in the remnant, which is determined by the properties of the medium. In particular, when the remnant expands in an inhomogeneous medium, the shock front penetrates between dense fragments, maintaining a higher velocity for a longer time \citep{Korolev2015,Slavin2017,Wang2018}. The evolution of dust grains located in diffuse gas and dense fragments differs \citep[e.g.,][]{Martinez2019,Kirchschlager2022,vs2024} and depends on the density dispersion in the medium \citep{Dedikov2024}. These papers discuss variations in the distribution of grain size in different thermal phases. However, no attention is paid to possible changes in the emission properties of interstellar dust swept-up by the SN shell. This paper is devoted to studying the influence of inhomogeneities in the surrounding medium on the IR emission of dust in the remnant. 

Section~2 describes the model and initial conditions. Section~3 presents the results. Section~4 discusses the application of the results and their consequenses. Section~5 summarizes the main conclusions.

\section{MODEL DESCRIPTION}

Let us consider the emission properties of dust and gas in the SN remnant expanding in an inhomogeneous medium. Using the numerical methods for the modelbing of the gas and dust dynamics described in \citep{vs2024}, we carry out a three-dimensional study of the evolution of an isolated SN remnant depending on the magnitude of inhomogeneity of the ambient medium and follow the dynamics of polydisperse interstellar dust in the swept-up SN shell \citep{Dedikov2024}. Here we thoroughly focus on the emission properties of dust. At first we give a brief description of the initial conditions, individual aspects of dust evolution in the SN remnant and the methods for calculating the emission of dust and gas.

\subsection{Initial Conditions}

To obtain an inhomogeneous gas density field, the pyFC \citep{Lewis2002}, is used, which allows generating “fractal cubes” with a lognormal amplitude distribution and a Kolmogorov spatial spectrum with index $\beta$ = 5/3. The density field is characterized by the mean value $\langle n\rangle$ and the standard deviation $\sigma$. In the models of the SN remnant evolution calculated by \citet{Dedikov2024}, the fiducial mean gas density is equal to $\langle n\rangle = 1$~cm$^{-3}$; the dispersion $\sigma$ varies from 0 to 3, which correspondes to a uniform gas distribution for $\sigma=0$ and a deviation of the density from the mean of up to 300 times for the maximum $\sigma$. In addition, we calculate several models with different mean densities $\langle n\rangle =  0.3, 3, 10$~cm$^{-3}$ for some values of $\sigma$. The maximum size of density fluctuations in the models considered here is 6~pc (it is determined by the number of cells along the side of the fractal cube $N_c$, the minimum value of the wave number $k_{min}$ and the spatial resolution of the cube $\Delta x$). 

Regardless of the density field variations, it is assumed that the gas is initially in thermal equilibrium, i.e. $\rho T$ = const, with zero velocities of the gas and dust particles. To account for radiation losses, we use  the nonequilibrium cooling function \citep{v11,v13}. It was obtained for the isochoric process of gas cooling from $10^8$~K до 10~K, including the ionization kinetics of all ionic states of the following chemical elements: H, He, C, N, O, Ne, Mg, Si и Fe. The rate of the gas heating is set to a value to stabilize the medium, not perturbed by the shock wave from the SN.

The dust-to-gas density ratio is taken to be 0.01, which is typical for the solar metallicity as it is assumed for the interstellar gas in our models \citep{Dedikov2024}. At the initial time, the grain size distribution follows a power law with index $-3.5$ \citep{Mathis1977} in the range $30-3000$~\AA, divided into 11 equal bins on a logarithmic scale. The minimum grain size in the calculations is 10~\AA. The destruction of dust due to sputtering is taken into account \citep{Draine1979a}. 

The characteristic time of destruction of dust grains due to shattering turns out to be longer than the calculation time. For instance, in a warm ionized medium with $(T,n)=(8\times10^3$~K,~0.1 cm$^{-3})$, the typical time for this process is more than $\sim 1-5$~Myrs, and in warm $(6\times10^3$K,~0.3~cm$^{-3})$ and cold $(10^2$~K,~30~cm$^{-3})$ neutral media it even increases to several tens of millions of years \citep{Hirashita2009}. In SN remnants, the collision time between grains varies in the range of $\sim 4-40$~Myrs \citep{Martinez2019}. Note that the growth of grains is most efficient in dense and cold environments, but for the conditions considered here this time is more than several tens of millions of years \citep{Zhukovska2008}. Therefore, these processes can be neglected.

When a supernova explodes, mass and energy are injected into a small region with size of 1.5 pc. There are four cells per radius of this region for the spatial resolution of 0.375 pc used here. The SN energy is $10^{51}$~erg, it is added as thermal energy. The masses of the injected gas and metals are 30 and 10~$\msun$, respectively.

For numerical solution of gas dynamics equations, an explicit scheme without splitting of fluxes on space and with the condition of total variation diminishing (TVD) is used, which provides high-resolution capturing of shocks and prevents unphysical oscillations The scheme belongs to the type of Monotonic Upstream-Centered Scheme for Conservation Laws (MUSCL-Hancock). To improve the accuracy of calculating fluxes at cell boundaries, the approximate Harten–Lax–van Leer-Contact (HLLC) method is used to solve the Riemann problem \citep[e.g.,][]{Toro2009}. 

The dust dynamics is described using the “superparticle” method proposed by \cite{Youdin2007}. A superparticle is a conglomerate of identical microparticles -- dust grains. For each superparticle, the equations of motion are solved taking into account the mutual influence on the gas due to friction forces \citep{Epstein1924,Baines1965,Draine1979a} and the equation for the radius of the dust particle due to thermal and kinetic sputtering \citep{Draine1979b}. The methods were considered in more detail by \citep{Mignone2019,Moseley2023} and adapted in the software package used here \citep[see the description and tests in Appendix A in][]{vs2024}.

\subsection{Evolution of the Remnant}

\begin{figure*}
\center
\includegraphics[width=15cm]{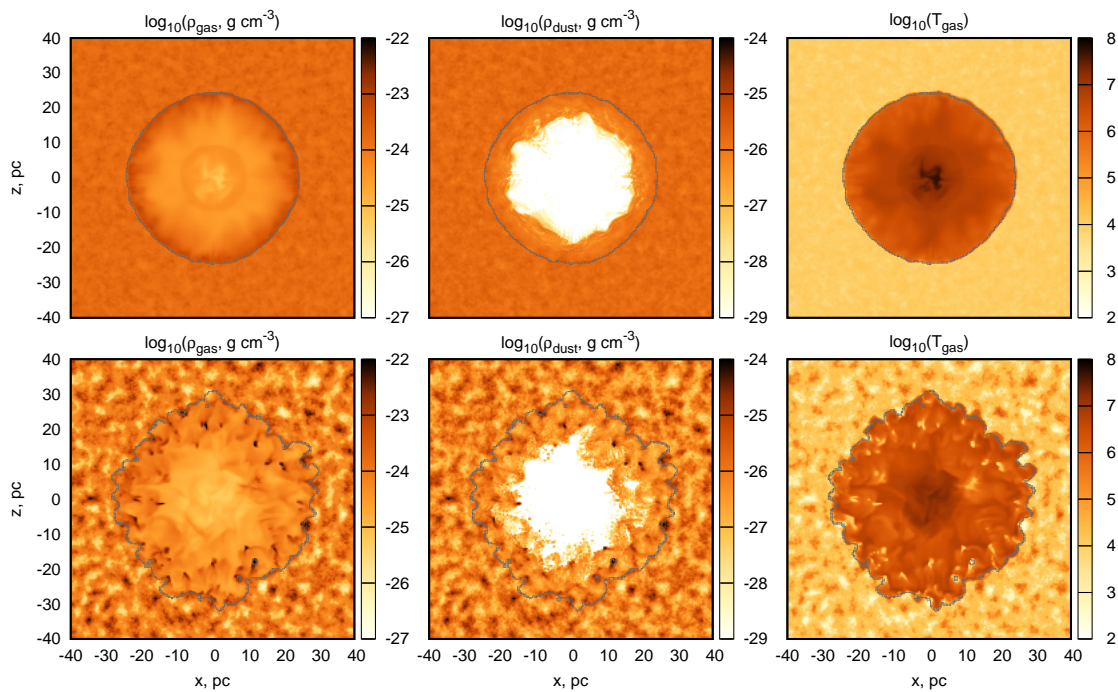}
\caption{
Distributions of gas density (left panels), interstellar dust density (middle panels) and gas temperature (right panels) in the plane passing through the center of the SN remnant expanding in an inhomogeneous medium with a lognormal distribution of density perturbations with mean density $\langle n \rangle = 1$~cm$^{-3}$ and variance $\sigma=0.2$ (upper row of panels) and 2.2 (lower row of panels) at the time of 40 kyr. The gray line corresponds to the outer boundary of the SN remnant, determined from the jump in gas velocity.
}
\label{fig-den-maps}
\end{figure*}

During several thousand years\footnote{This period includes free expansion phase and intermediate phase, during which the mass of the swept-up shell becomes significantly greater than the mass of the gas ejected by the SN.} after the explosion, the SN shell sweeps interstellar gas containing dust. Depending on the background density of the medium, the SN shell goes either to the adiabatic (at a density of $\simlt 1$~cm$^{-3}$) or immediately to the radiative (at $\simgt 10$~cm$^{-3}$) expansion phase. During expansion in an inhomogeneous medium, the SN shell interacts with gas of different densities, and the shock wave penetrates into regions of lower density at a higher velocity and, conversely, slows down in dense clouds \citep{Korolev2015,Slavin2017,Wang2018}. Figure~\ref{fig-den-maps} shows the gas density distributions (left and middle columns of the panels) for a SN remnant of age 40~kyr old expanding in a medium with low ($\sigma=0.2$, see upper panels of Figure~\ref{fig-den-maps}) and high ($\sigma=2.2$, lower panels of Figure~\ref{fig-den-maps}) density fluctuations. The features of the SN remnant evolution are described in detail in \citep{Dedikov2024}. Note that, due to its inertia interstellar dust penetrates far behind the shock front (shown by the grey line) and enters the gas with $T\simgt 10^6$K и $n\simlt 0.1$~cm$^{-3}$ (Figure~\ref{fig-den-maps}), in which favorable conditions exist for both grain sputtering and their efficient emission.

\subsection{Dust and gas Emission}

Heating of dust grains in hot gas of the SN remnant is produced mainly by collisions with electrons \citep{Draine1979a}. Small particles ($a \sim 30$~\AA ) experience strong temperature fluctuations in hot gas, since the characteristic cooling time of a grain is comparable or shorter than the average time between collisions. Therefore, the stochastic method is used to calculate the temperature of such grains \citep{Draine1985}. Computing the temperature distribution functions (TDF) of dust grains\footnote{{Here we mean the temperature of the matter of a grain. It is necessary to distinguish the temperature of the grain matter from the dust temperature determined from the maximum of the (modified) Planck spectrum, which is an effective (averaged) value over the emission spectra of dust particles. Usually these quantities are referred to by the same term: “dust temperature” \citep[see, for example,][]{Dwek1992}.} } is based on direct simulation of particle--grain collisions using the Monte Carlo method (a detailed description of the implementation of the TDF computation procedure can be found in \citep{Drozdov2021}. At the same time, for grains with sizes $\simgt 1000$~\AA\ the cooling time exceeds the interval between subsequent collisions; under these conditions, the dust temperature is close to the equilibrium value $T_{eq}$, which follows from the equality of energy losses due to radiation and heating during collisions: $L_{IR}(a, T_{eq}) = H_{coll}(a,T_{g} ,n_{e})$, where $a$ is the grain size, $T_g$ is the gas temperature, and $n_e$ is the electron density \citep[e.g.,][]{Dwek1992}. The contribution from ultraviolet (UV) photons of the diffuse background with energy density of the order of magnitude in the local interstellar medium \citep{Habing1968} turns out to be insignificant during the period of the remnant evolution considered here and can be neglected. In this work, we are more interested in the emission properties of dust in the mid- and far-IR ranges, so the spectral features associated with elastic vibrations in the molecular structure of small grains -- polyaromatic hydrocarbons (PAHs) -- with sizes less than $30$~\AA \ and being significant at short wavelengths are not taken into account.

To speed up the calculation of the heating rate of dust grains, a grid of the TDFs has been calculated for a wide range of gas parameters: $T = [4\times 10^4 - 10^8]$~K, $n = [10^{-5} - 10^2]$~cm$^{-3}$. Using the obtained distribution functions, we calculate the emission spectra of dust grains with sizes $30-3000$~\AA, divided into 11 equal intervals on a logarithmic scale. For grains in a gas with $T \sim 10^4 - 4\times 10^4$~K, the emission spectra is obtained for the equilibrium value of $T_{eq}$. The optical parameters of dust grains are taken from  \citep{Draine1984,Laor1993}. The dust is assumed to be composed of silicate and graphite particles with equal mass fractions {\citep[e.g.,][]{Yamada2005,Corrales2016}, although other ratios are also used \citep[e.g.,][]{Draine1984}}. The emission spectrum is calculated in the wavelength range from 1 to 1000~$\mu$m. Thus, to construct dust emission maps, we sum the spectra from all dust grains affected by local physical conditions in each of the numerical cells along the line of sight.

To estimate the dust temperature$^{2)}$ a modified Planck spectrum with absorption index $\beta = 2$ {\citep[for grains without icy mantles, e.g.,][]{Draine1984}}, the temperature of the maximum of which can be determined as follows \citep[see, e.g., equation I.76 in][]{Galliano2022}:
\begin{equation}
 T_{d} = \frac{hc}{k_{B} \lambda_{max}}\frac{1}{(4 + \beta) + W[-(4+\beta){\rm exp}(-(4 + \beta))]}
\label{Twien}
\end{equation}
where $W[..]$ is the Lambert W--function, $\lambda_{max}$ is the wavelength of the maximum of the spectral luminosity. 

\begin{figure*}
\center
\includegraphics[width=12.5cm]{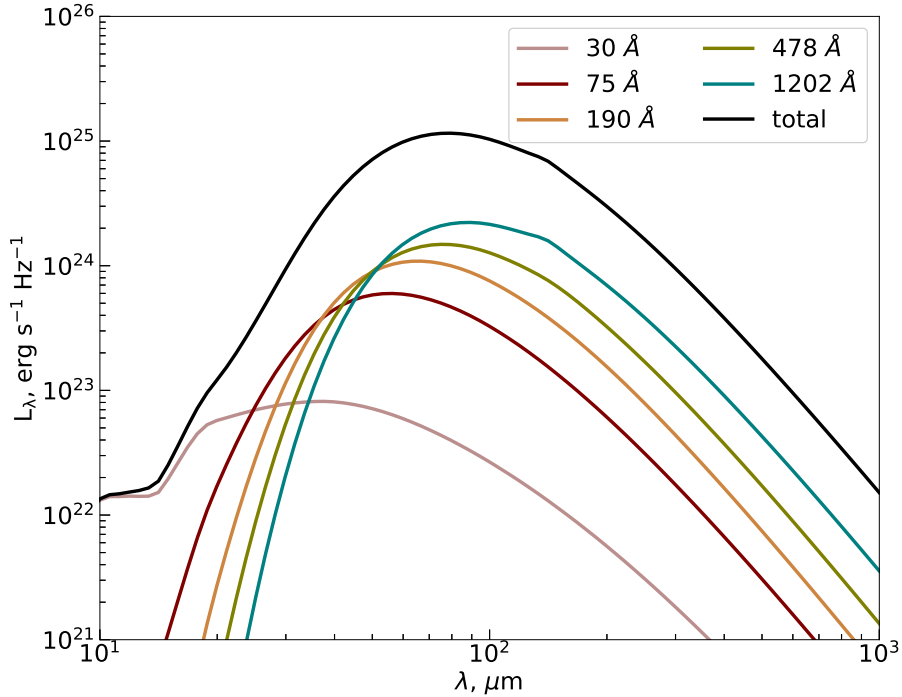}
\caption{
IR spectra for SN remnant with age of 40~kyr evolving in a medium with a lognormal distribution of density perturbations with mean value $\langle n \rangle = 1$~см$^{-3}$ and variance $\sigma=2.2$. The black line represents the total spectrum, the colored lines are spectra of dust grains with sizes in several intervals (the value in the legend corresponds to the center of the size interval).
}
\label{fig-exmplspec}
\end{figure*}

Figure~\ref{fig-exmplspec} shows the IR spectra of the SN remnant with age of 40~kyr expanding in a medium with mean density $\langle n \rangle = 1$~см$^{-3}$ and dispersion $\sigma = 2.2$. The spectra are calculated in the range 1--1000~$\mu$m divided by 120 bins on a logarithmic scale. The dust temperature varies from 20 to 80 K, which corresponds to $\sim$ $45-175$~$\mu$m, this interval accounts for 24 bins. The main contribution to the luminosity in the region of the spectrum maximum ($\lambda \simeq 100$~$\mu$m) is made by large grains with size of $a \simgt 500$~\AA, in the shortwave range ($\lambda < 40$~$\mu$m) the emission from small grains with $a \simlt 200$~\AA, dominates. Contributions from small grains extend the spectrum to shorter wavelengths and apparently shift its maximum. According to equation (\ref{Twien}), the temperature of large grains with size of $\sim 1200$~\AA \ is about 40~K, small grains with size of $\sim 75$~\AA \ are “heated” to 60~K, for $a \sim 30$~\AA \ -- almost to 80~K. In this case, the dust temperature for the total spectrum turns out to be 40~K, which corresponds to the typical value for large grains. During the period of SN evolution considered here, the picture remains the same: the dust temperature is determined by large particles with size of $\simgt 1000$~\AA. Note that at wavelengths shorter than 20~$\mu$m the contribution of PAHs to the total IR luminosity of dust does not exceed a few percent for typical values of the PAH content $q_{\rm PAH} \sim 1$\% and the flux of external UV radiation in the local interstellar medium $U \sim 1$ \citep[][]{Draine2007}.

To calculate the luminosity in metal lines, the emissivity of each gas element (grid cell) is calculated, which is determined by the gas temperature and the number density of the corresponding ionic state; the value of the latter at a given temperature is found by using the pre-computed tables for nonequilibrium cooling of a gas \citep{v13}. Afterwards, the emission in the line is integrated along each line of sight.

\section{RESULTS}

\begin{figure*}
\center
\hspace{1.0cm}
\includegraphics[width=12.5cm]{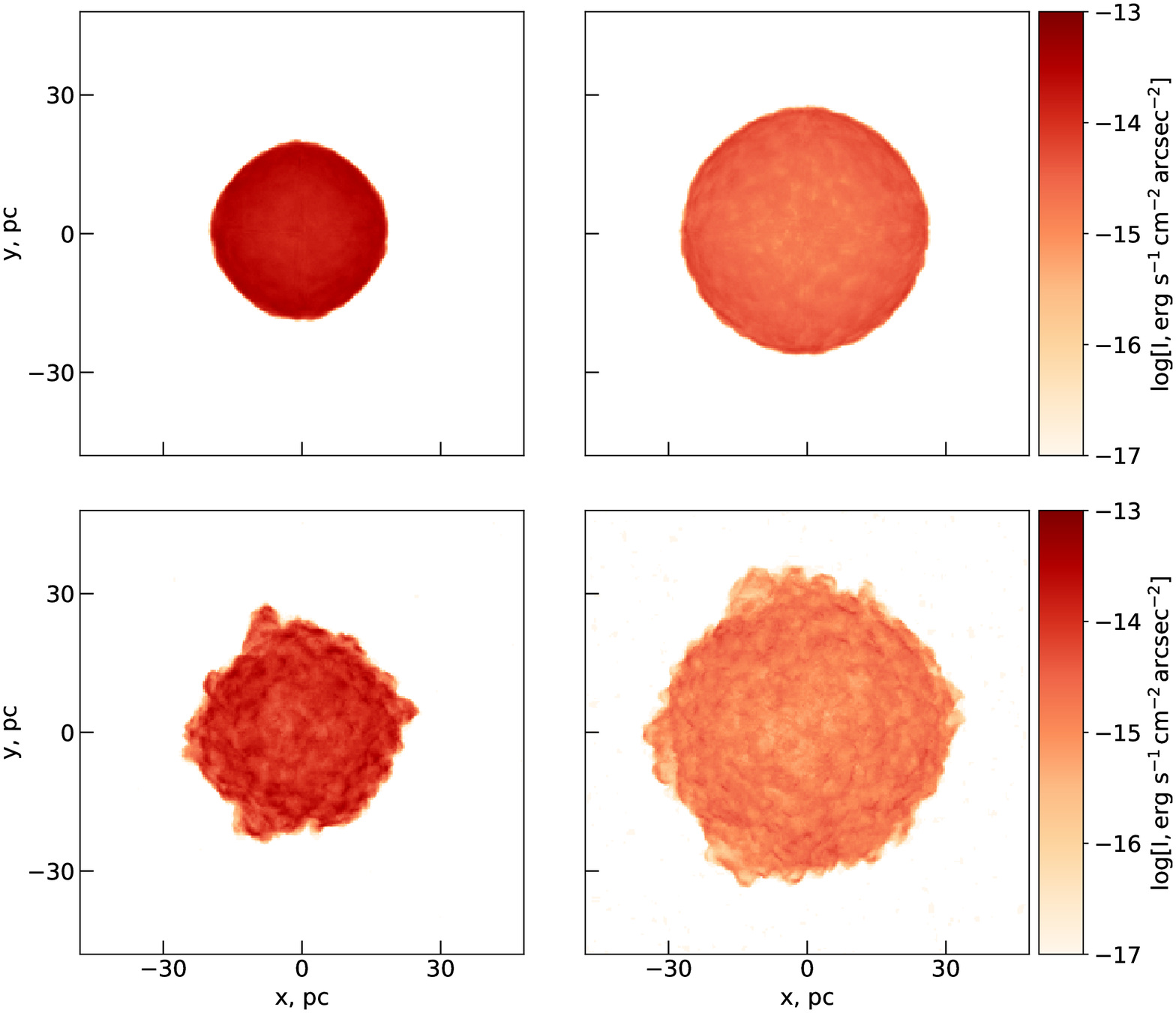}
\includegraphics[width=11.5cm]{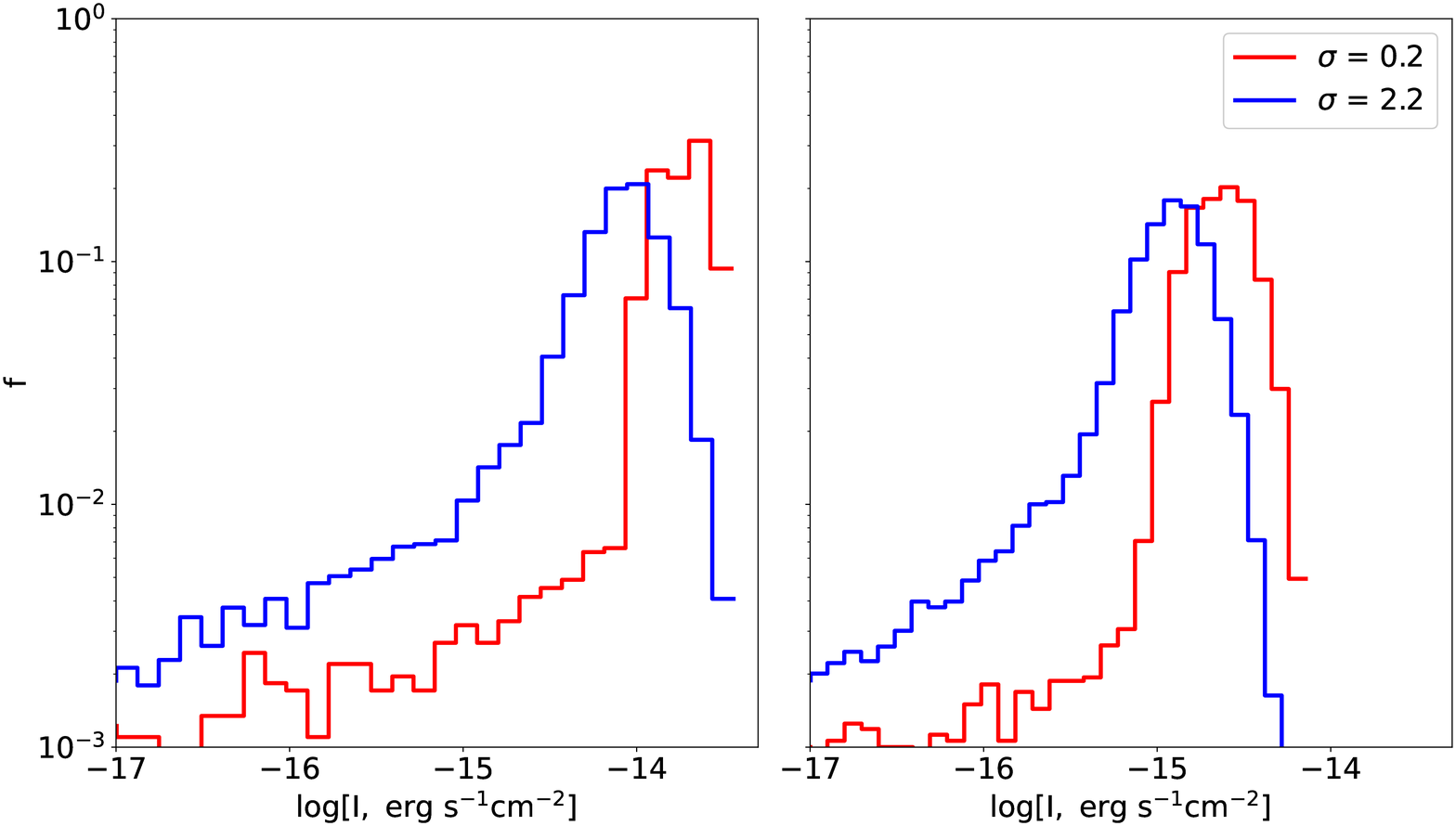}
\caption{
The IR surface brightness maps for the SN shell expanding in an inhomogeneous medium with density dispersion $\sigma$ = 0.2 (upper panels), 2.2 (middle panels), and the corresponding distributions of surface IR brightness (two bottom panels) for the maps shown above at 20~kyr (two upper left panels) and 50 kyr (two upper right panels).) 
}
\label{fig-ir-maps}
\end{figure*}

Let us calculate the dust emission from an SN remnant expanding in an inhomogeneous medium with density dispersion $\sigma$. Figure~\ref{fig-ir-maps} shows the surface brightness maps (along the line of sight, with summing up along the $z$ axis and in the wavelength range from 1 до 1000~$\mu$m) in the IR range from dust in the remnant with age of 20 and 50 kyr, expanding in a medium with weak ($\sigma= 0.2$) and high ($\sigma= 2.2$) density fluctuations. It should be noted that the surface brightness is distributed almost uniformly, which is expected, since the dust in the remnant is located in a thick shell without any noticeable fluctuations (Figure~\ref{fig-den-maps}). It can be seen that the surface brightness of the remnant decreases in time and with increasing magnitude of inhomogeneity of the medium {(see Figure~\ref{fig-ir-maps})}. The former is associated with cooling of the gas, the latter appears owing to higher fraction of dense and cold regions (fragments) that the SN shock wave is unable to destroy and heat. 

\begin{figure*}
\center
\includegraphics[width=16cm]{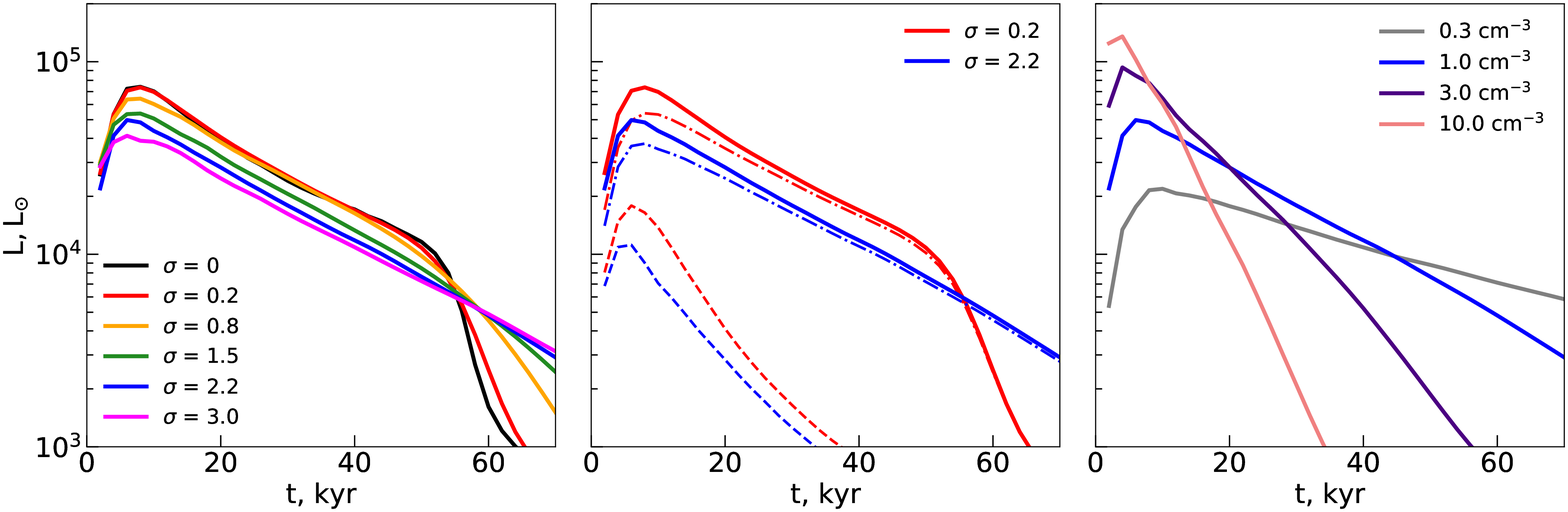}
\caption{
The evolution of the total IR dust luminosity (in $\lsun$) in the SN remnant expanding in an inhomogeneous medium for different values of the density dispersion $\sigma$ and mean density $\langle n \rangle = 1$~cm$^{-3}$ (left panel). The total IR luminosity is shown by solid lines. The contributions to the total IR dust luminosity in the ranges $1-30$~$\mu$m (dashed lines) and $30-1000$~$\mu$m (dash-dotted lines) in the remnant expanding in a medium with $\sigma = 0.2; 2.2$ and $\langle n \rangle = 1$~cm$^{-3}$ (middle panel). The dependence of the IR luminosity for different values of mean gas density $\langle n \rangle$ and dispersion $\sigma = 2.2$ (right panel).
} 
\label{fig-evol-lum}
\end{figure*}

\begin{figure*}
\center
\includegraphics[width=14.5cm]{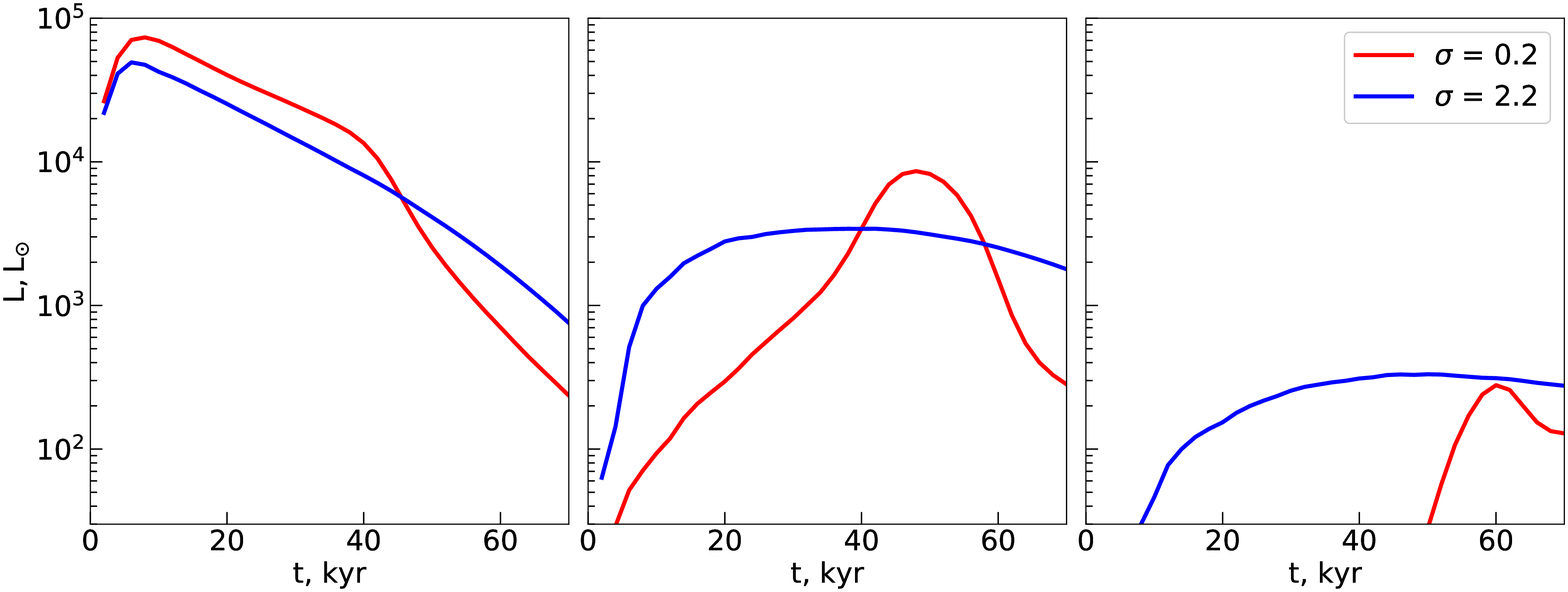}
\caption{
The IR luminosity of dust located in a gas with $T>10^6$K (left panel), $10^5<T<10^6$K (middle) and $10^4<T<10^5$K (right) in the SN remnant expanding in an inhomogeneous medium with gas density dispersion $\sigma$ = 0.2 (red solid line) and 2.2 (blue solid line). The mean gas density is $\langle n \rangle = 1$~cm$^{-3}$. 
}
\label{fig-lum-tphases}
\end{figure*}

Figure~\ref{fig-evol-lum} shows the evolution of the total (in the wavelength range 1--1000~$\mu$m) IR luminosity of the SN remnant evolving in a medium with different magnitudes of inhomogeneity, i.e., density dispersion $\sigma$. Figure~\ref{fig-evol-lum} (middle panel) shows the contributions to the total IR luminosity of dust in the ranges $1-30$~$\mu$m (dashed lines) and $30-1000$~$\mu$m (dash-dotted lines) in the SN remnant. It is evident that the luminosity in the longwave range dominates during the period of the SN evolution considered here.
In the first few thousand years, the swept-up interstellar dust penetrating behind the shock front is effectively heated and destroyed, since the gas temperature behind the front reaches tens of millions of K.
The mass of the swept-up dust grows rapidly until the onset of the adiabatic phase, after which the rate of interstellar dust inflow decreases. Around this time moment, the IR luminosity of the remnant reaches its maximum: $L\sim (4-8)\times 10^4\lsun$. Afterwards this value is reduced due to the decrease in gas temperature and by the age of the remnant of about 50 kyr it is about $\sim 10^4\lsun$. In gas with $T\simlt 10^5$K, the efficiency of heating dust particles drops significantly, and therefore the IR luminosity of the remnant decreases catastrophically below $10^3\lsun$ after the greater mass fraction of the SN shell turns out to be colder than $10^5$K. For the remnant expanding in a lower inhomogeneous medium with $\sigma\simlt 0.2$, the thermal phases of the shell are pronounced in time: $T\simgt 10^6$~K -- up to 40 kyr, that is, before the onset of the radiative phase, $T\sim 10^5 - 10^6$~K -- within 40 -- 60 kyr, $T\simlt 10^5$~K -- after 60 kyr. It is clearly seen in Figure~\ref{fig-lum-tphases} (left panel) that the main contribution to the IR luminosity during SN evolution in a medium with $\sigma = 0.2$ is given by dust associated with $T\simgt 10^6$K up to the SN age of about 45~kyr. Then, dust associated with gas of $T\sim 10^5 - 10^6$K dominates, its radiation rapidly decreases after 60 kyr (Figure~\ref{fig-lum-tphases}, middle panel). Dust in colder gas is heated slightly, and its contribution to the total luminosity is insignificant (Figure~\ref{fig-lum-tphases}, right panel).

In more inhomogeneous medium, various thermal phases of gas, with which dust is associated, can be found inside the remnant at almost any time after the first few thousand years. This is due to that the shockwave penetrates into regions with lower density at higher velocity and, conversely, slows down in dense clouds \citep{Korolev2015,Slavin2017,Wang2018}. For this reason, when expanding in inhomogeneous media, gas with $T\simgt 10^5$~K is preserved inside the remnant for much longer. In more inhomogeneous media with $\sigma \simgt 2.2$ the IR luminosity of the remnant decreases at a constant rate during the whole period of the SN evolution considered here (Figure~\ref{fig-evol-lum}) and by the age of about 70~kyr it turns out to be $\sim 3\times 10^3\lsun$, which is almost an order of magnitude higher than in the case of evolution in lower inhomogeneous medium. In Figure~\ref{fig-lum-tphases} (left panel) one can see that the IR luminosity of dust located in hot gas ($T\simgt 10^6$K) behaves similarly. The contribution from dust located in warm gas ($T\sim 10^5 - 10^6$~K) remains almost at the same level after 10~kyr, which after 60~kyr turns out to be dominant (Figure~\ref{fig-lum-tphases}, middle panel). An increase of the mean density of the medium in which the SN remnant expands leads to growth of the maximum IR luminosity, earlier reaching it, and significantly faster decrease with time (Figure~\ref{fig-evol-lum}, middle panel). Thus, the IR luminosity depends significantly on the magnitude of inhomogeneity and the mean density of the gas in which the SN remnant expands.

\begin{figure*}
\center
\includegraphics[width=16cm]{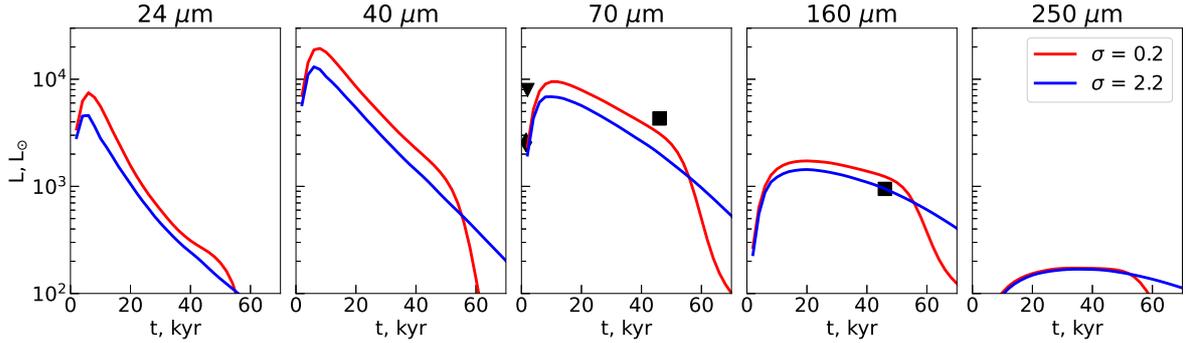}
\caption{
The evolution of dust luminosity in spectral bands 24, 40, 70, 160 and 250~$\mu$m (panels from left to right) for an SN remnant expanding in an inhomogeneous medium with gas density dispersion $\sigma$ = 0.2 (red solid line) and 2.2 (blue solid line). The bandwidth is $\Delta \lambda = \lambda/3$. The symbols indicate the observed luminosity of several SN remnants: G304.6 \citep[square,][]{Lee2011}, 3C~397 \citep[diamond,][]{Koo2016}, G34.7 \citep[triangle,][]{Koo2016}.
}
\label{fig-evol-lum-bands}
\end{figure*}

In addition to gas cooling, the decrease in the IR luminosity is affected by sputtering of dust grains during collisions with protons in the hot gas. Smaller particles are more susceptible to destruction: the characteristic lifetime of particles smaller than 100\AA \ in gas with T $\simgt 10^6$~K and density $\sim 0.1$~cm$^{-3}$ does not exceed several tens of thousands of years \citep{Draine1979b}. Thus, by the onset of the radiative phase a significant part of the small dust in the SN shell will has been destroyed \citep[see Figures~5 and 9 in][]{Dedikov2024}. In the first few thousand years of the SN evolution, its IR luminosity increases due to the heating of small dust particles $a\simlt 100$\AA. The maximum IR luminosity of the SN remnant, reached by this age, is determined by the dust emission in the spectral band centered at 40 $\mu$m (Figure~\ref{fig-evol-lum-bands}, second panel). The fraction of shortwave radiation coming in the 24 $\mu$m, band is up to 20 -- 30\% of the total IR emission of the remnant in the epoch of 5 -- 15~kyr (Figure~\ref{fig-evol-lum-bands}, first panel). Gradually, these grains are destroyed and the main role in IR emission passes to larger particles with $a\sim 200-1000$~\AA \ (Figure~\ref{fig-exmplspec}). Although smaller dust particles continue to arrive behind the shock front due to the expansion of the remnant, their heating efficiency decreases, and they no longer make a noticeable contribution to the total luminosity. During the period from 20~kyr to almost 60~kyr the main part of the radiation comes in the 70~$\mu$m band (Figure~\ref{fig-evol-lum-bands}, third panel). Only at the late evolutionary phases, after 60~kyr, a significant contribution to the dust emission comes in the 160~$\mu$m band (Figure~\ref{fig-evol-lum-bands}, forth panel). The radiation in the most longwave band (250~$\mu$m) turns out to be insignificant during the whole period of the SN remnant evolution considered here (Figure~\ref{fig-evol-lum-bands}, fifth panel). Therefore, the 70~$\mu$m band can be considered as the most optimal range for tracking the evolution of the dust emission properties in the SN remnant. In the vicinity of this wavelength there are important indicators of the thermal state of the gas in the SN shell -- the oxygen and nitrogen lines: [OIII]~52~$\mu$m, [NIII]~57~$\mu$m, [OI]~63~$\mu$m, [OIII]~88~$\mu$m. In particular, for the latter we consider below the dependence of its luminosity on the properties of the medium, where the SN remnant is expanding, and the ratio of its luminosity to the dust continuum under the line.

The measurements of dust IR emission in the 70 $\mu$m and 160~$\mu$m bands from the remnants of G304.6 \citep[shown by square symbol in Figure~\ref{fig-evol-lum-bands},][]{Lee2011}, 3C~397 \citep[diamond,][]{Koo2016}, G34.7 \citep[triangle,][]{Koo2016} are close to the values calculated in the models. Note that among several dozen observations of SNe remnants in the IR range \citep[][и др.]{Pinheiro2011,Chawner2020}, most of them lack age estimates or do not have measurements in the required bands. Note that for a significant number of young (1 -- 2~kyr) SN remnants, the age is determined more accurately \citep[e.g.,][]{Koo2016,Chawner2020}, therefore, these data could be compared with the presented models. However, firstly, at such a small age the radiation from the dust produces by the remnant can still be significant and our model does not include the dynamics of such dust. At late times, the mass of the swept-up interstellar dust significantly exceeds the mass produced by the remnant. Secondly, in young remnants the dust is apparently quite hot \citep[e.g.,][]{Koo2016,Priestley2022} and radiates predominantly in the shortwave range. Therefore, the interpretation of observations of young remnants requires specific numerical calculations, in particular, taking into account the evolution of injected dust particles and higher spatial resolution.

\begin{figure*}
\center
\includegraphics[width=16cm]{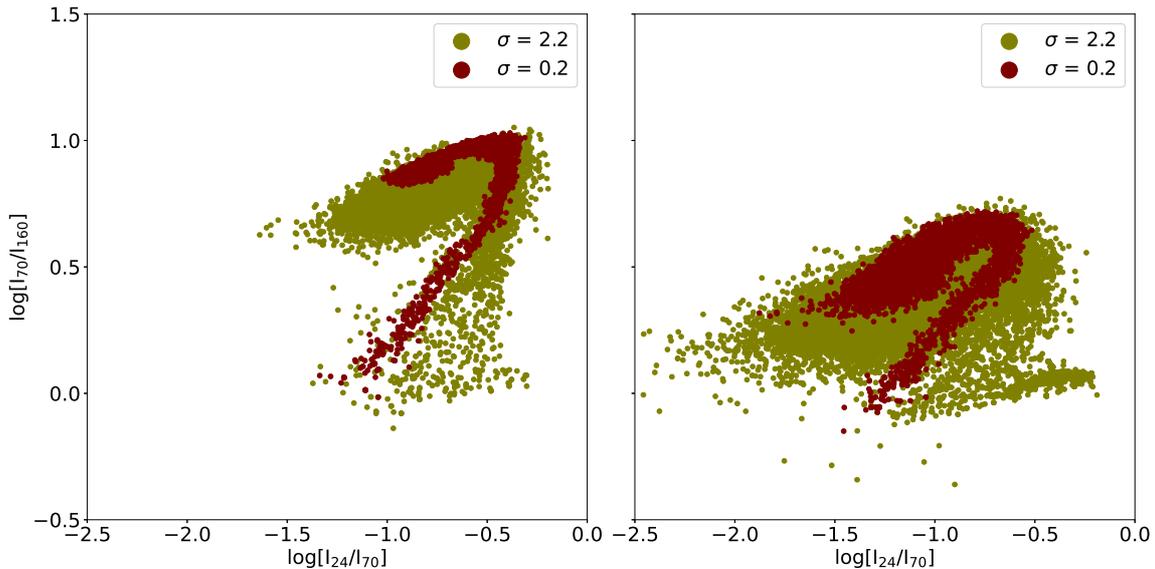}
\caption{
The ratio of surface brightness values in the bands for the distribution of IR luminosity of dust in the SN shell expanding in an inhomogeneous medium with density dispersion $\sigma$ = 0.2 (dark symbols), 2.2 (сlight symbols), at times of 20 (left panel) and 50~kyr (right panel).
}
\label{fig-evol-ratio-bands}
\end{figure*}

The surface distribution of IR luminosity over the SN remnant is almost flat (Figure~\ref{fig-ir-maps}), small fluctuations arise due to inhomogeneities of the medium through which the shell propagates. The distribution of surface IR brightness of the SN remnant in spectral bands should behave similarly. In the color diagrams (Figure~\ref{fig-evol-ratio-bands}) for the ratio of surface brightness values in the bands $I_{24}/I_{70}$ and {$I_{70}/I_{160}$} (the subscript is the central wavelength of the band) one can see that during evolution in more inhomogeneous medium the spread of values increases. This is more clear for young remnant at adiabatic  phase (Figure~\ref{fig-evol-ratio-bands}, left panel, see the middle region with light and dark symbols). For older remnant (Figure~\ref{fig-evol-ratio-bands} right panel) the average values decrease, the dispersion of the ratios increases slightly. As the age of the remnant increases, the luminosities in the shortwave bands decrease more sharply and the locus of points shifts noticeably on the diagram. This difference in the distribution of values on the color diagram may indicate the magnitude of inhomogeneity of the medium in which the remnant evolves.

\begin{figure*}
\center
\includegraphics[width=14.2cm]{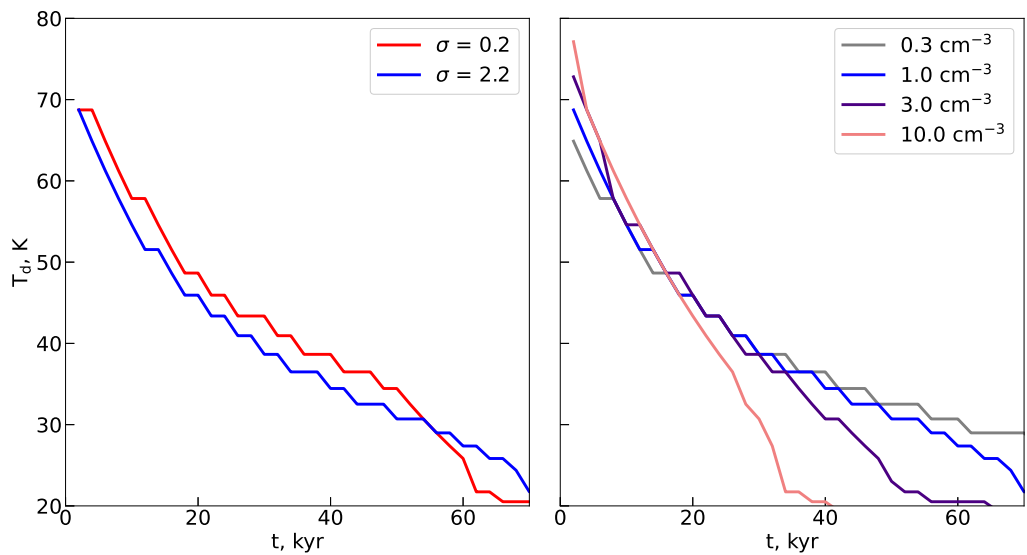}
\caption{
The dust temperature determined from the maximum of the IR emission spectrum of an SN remnant expanding in an inhomogeneous medium with dispersion $\sigma$ = 0.2 and $\sigma$ = 2.2 (red and blue solid lines, respectively) in the case of an average gas density $\langle n \rangle = 1$~cm$^{-3}$ (left panel) and $\sigma$ = 2.2 for several values $\langle n \rangle = 0.3$, 1, 3 и 10 cm$^{-3}$ (right panel).
}
\label{fig-evol-dust-temp}
\end{figure*}

The shift of the wavelength band, in which the IR emission of the remnant mainly contributes, indicates a change in the dust temperature. Using spectral distributions similar to that shown in Figure~\ref{fig-exmplspec}, we obtain the evolution of the average dust temperature in the SN remnant for each time moment. Figure~\ref{fig-evol-dust-temp} shows the dependence of this value for density dispersion $\sigma=0.2$ and 2.2, corresponding to lower and more inhomogeneous media (Figure~\ref{fig-evol-dust-temp}, left panel), and for several mean density values in the ambient medium $\langle n \rangle$ at fixed dispersion $\sigma=2.2$ (Figure~\ref{fig-evol-dust-temp}, right panel). Flat parts on the curves appear due to the relatively low resolution of the calculated spectra (see, for example, Figure~\ref{fig-exmplspec}) and the slow shift of the maximum in the spectrum during the evolution of the remnant. At first, one can see that the dust temperature drops from $T_d \sim 70$~K in the first 10~kyr to $\simlt 30$~K by the age of 60~kyr. These values are significantly higher than the typical dust temperature of about $\sim 20$~K in the medium not affected by the SN shock. At second, the difference between the models with $\sigma=0.2$ and $\sigma=2.2$ turns out to be insignificant during the evolution and is about 5~K (Figure~\ref{fig-evol-dust-temp}, left panel), which can hardly be used to determine the magnitude of inhomogeneity of the medium in which the SN remnant evolves. Further, at late phases of the evolution, one can note the dependence of the dust temperature on  mean density of the medium (Figure~\ref{fig-evol-dust-temp}, right panel). For $\langle n\rangle \sim 10$~cm$^{-3}$ the value of $T_d$ in the remnant with age of about 40~kyr drops almost twofold compared to the value for 1~cm$^{-3}$ and turns out to be close to the background value, equal $\sim 20$~K. The decrease in the average density of the medium leads to reaching the background value at a later time.

\begin{figure*}
\center
\includegraphics[width=14.2cm]{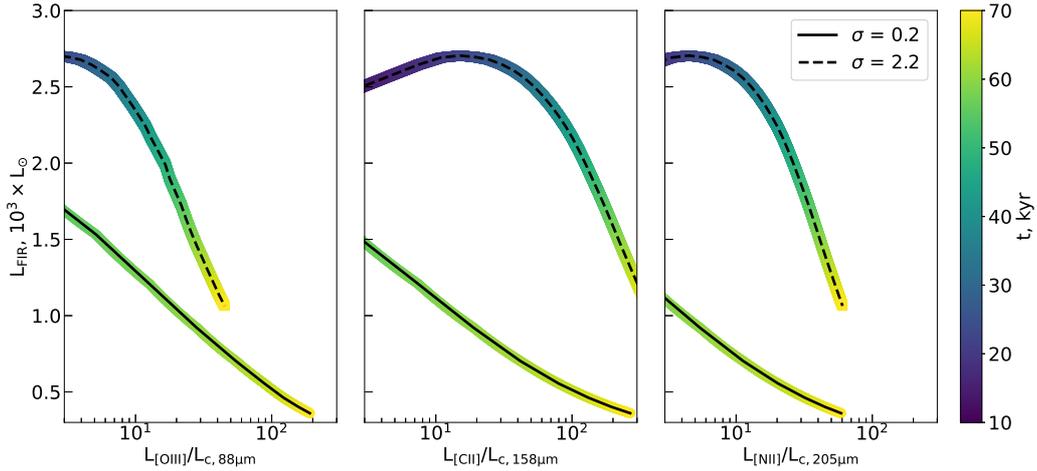}
\caption{
The dependence of the ratio of luminosity in several spectral lines $L_i$ to the luminosity of the dust continuum under the line in the 8~GHz band $L_{c,i}$ on the luminosity in the far-IR range $L_{\rm FIR}$ ($\lambda = 100-1000$~$\mu$m). The color bar shows the age of the remnant in thousands of years. The solid curve displays the dependence for the remnant evolving in a medium with gas density dispersion $\sigma$ = 0.2, the dashed curve is for $\sigma$ = 2.2. The mean gas density is 1~cm$^{-3}$.
}
\label{fig-ratio-fir-lines}
\end{figure*}

After the onset of radiative cooling in the massive shell of the SN remnant, strong metal ion lines appear against the dust continuum. Figure~\ref{fig-ratio-fir-lines} shows the evolution of the ratio of the luminosities in the lines [OIII]~88~$\mu$m, [CII]~158~$\mu$m, [NII]~205~$\mu$m and in the continuum below the line in the band with width of 8~GHz depending on the total dust luminosity in the far-IR range $\lambda = 100-1000$~$\mu$m. As mentioned above, the onset of the radiative phase depends on the density dispersion $\sigma$ in the surrounding gas. Therefore, in a lower inhomogeneous medium ($\sigma = 0.2$) the lines become sufficiently strong (the luminosity in the line exceeds the underlying continuum) only in the SN shells with age of over 50 kyr. Increasing density dispersion in the medium the radiative phase begins earlier: for $\sigma=2.2$ strong metal lines appear already at age of about~30 kyr. A significant part of the line radiation comes from undestroyed fragments behind the shock front. In Figure~\ref{fig-ratio-fir-lines} the evolutionary diagrams for two values of $\sigma$ differ significantly, therefore, they can be used to determine the properties of the medium in which the SN remnant expands.

\begin{figure*}
\center
\includegraphics[width=12.5cm]{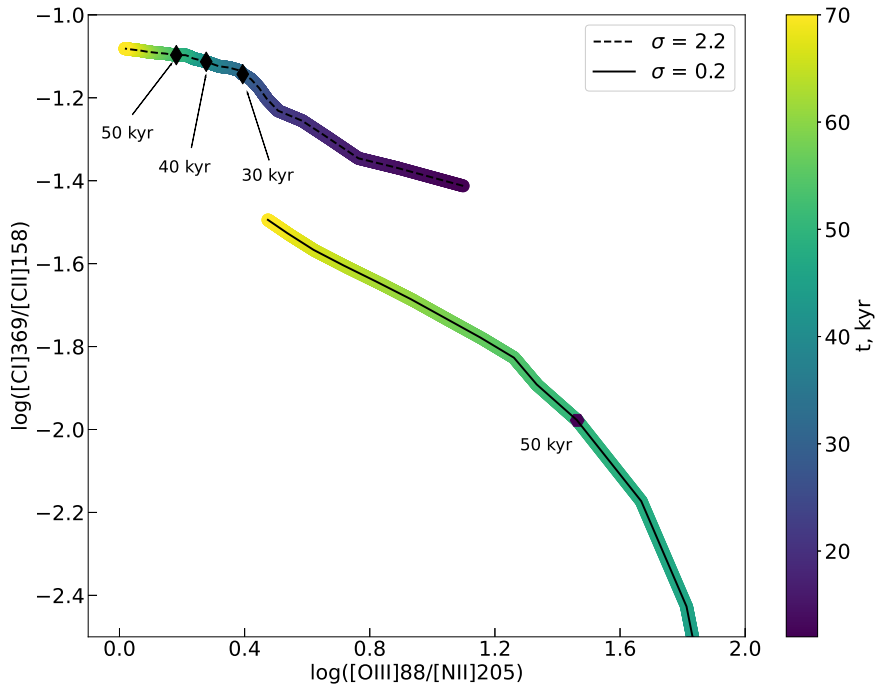}
\caption{
The ratios of the luminosities in the spectral lines of metal ions. The color bar corresponds to the age of the remnant in thousands of years. The solid line shows the ratio for the remnant expanding in a medium with the gas density dispersion $\sigma$ = 0.2, the dashed line is for $\sigma$ = 2.2. The mean gas density is 1~cm$^{-3}$. The labels mark different time moments.
}
\label{fig-ratio-lines}
\end{figure*}

Metal lines and their ratios can be used to determine the properties of the medium in which the remnant is expanding. Figure~\ref{fig-ratio-lines} clearly shows that the ranges of the ratios of the line luminosities differ significantly for remnants evolving in lower and more inhomogeneous gas. For intermediate values of the gas density dispersion, the evolutionary tracks lie between the curves shown in the diagram.

\section{DISCUSSION}

The IR emission of SN remnants older than several thousand years is apparently determined by the properties of the swept-up dust. A part of such dust is destroyed during further 10 -- 40~kyr. According to the calculations by \citet{Dedikov2024} the dust mass in the shell is $\sim 2-20~\msun$ depending on the gas density dispersion in a medium with a lognormal distribution and mean value of $\langle n \rangle \sim 1$~cm$^{-3}$. By the end of this period, almost $10~\msun$ of interstellar dust has been destroyed in the SN remnant. It is worth noting that in a more inhomogeneous medium, the mass inside the remnant is higher, since the shock front covers dense, partly destroyed fragments, inside which the dust is not subjected to sputtering. Obviously, an increase of gas density leads to growth of the swept-up dust mass. 

According to estimates by \cite{Chawner2020}, the dust mass in several SN remnants reaches tens and hundreds of solar masses (see Table 3 in \cite{Chawner2020}). Using the IRAS survey \citep{Arendt1989,Saken1992} even higher dust mass values were obtained for the objects G43.3--0.2 and G349.7+0.2. Dust temperature variations significantly reduce these values, but they still remain high. Such values can only be explained by the assumption that this is interstellar dust swept-up by the expanding SN shell. Moreover, if the remnant evolves in a denser and more inhomogeneous medium, the dust mass increases. In particular, even for a 20~kyr old SN remnant expanding in a medium with a lognormal gas density distribution for $\langle n \rangle \sim 10$~cm$^{-3}$ and $\sigma=2.2$, the mass of the swept-up dust will increase by about a factor of two, and one can expect that in  denser medium with $\langle n \rangle \sim 10$~cm$^{-3}$ the dust mass reaches $\sim 20-25~\msun$, which is comparable with the estimates of \citet[][see Table 3]{Chawner2020}. During the subsequent evolution, due to the deceleration of the SN shell in a dense medium, the dust mass increases and saturates at level of $\sim 40-45~\msun$. Apparently, the discrepancy between the dust mass estimates obtained for some SN remnants and the values following from the numerical models is associated with the strong sensitivity of observational quantities to small variations in dust temperature and the significant contribution of the interstellar dust.

The problem of the balance between dust production and destruction in the interstellar medium remains unresolved and quite serious \citep[e.g.,][]{Draine2009,Mattsson2021,Kirchschlager2022,Peroux2023}. A study of IR emission in SNe remnants older than 10 -- 20~kyr will shed light on the efficiency of dust destruction in media with different physical conditions. Some results in this direction have been obtained using the IRAS \citep{Arendt1989,Saken1992}, ISO \citep{Millard2021}, AKARI \citep{Ita2008,Kato2012} Spitzer \citep{Pinheiro2011,Seok2013,Matsuura2022}, Herschel \citep{Chawner2019,Chawner2020} space telescopes. However, the number of studied late SN remnants remains small. Therefore, it is desirable that the planned far-IR projects study such objects in more detail.

The surface brightness of the SN remnant in the continuum in the 8~GHz band at a frequency of 2~THz ($\lambda \simeq158$~$\mu$m) turns out to be on average about 10~MJy sr$^{-1}$ for ages from 20 to 60 kyr. During this period, the flux in the [CII] 158~$\mu$m line depends significantly on the density dispersion in the inhomogeneous medium (Figures~\ref{fig-ratio-fir-lines}--\ref{fig-ratio-lines}). This is due to the dynamics of the remnant expansion, how the shell cools and goes around dense dusty gaseous fragments, and how they are enclosed inside the remnant (Figure~\ref{fig-den-maps}, lower panels). The luminosity in the [CII] 158~$\mu$m line rapidly increases after the onset of the shell cooling and the cease of effective destruction of fragments, and exceeds the continuum by approximately a factor of 100 or more (Figure~\ref{fig-ratio-fir-lines}) and afterwards it remains at this level. Thus, for $\sigma\sim 0.2$ the luminosity increases to such high values for a remnant with age of about 60~kyr and older, and for $\sigma \sim 2.2$ -- already since 40~kyr. The intensity of this line after saturation reaches about $\sim 2\times 10^{-16}$~W/(m$^2 \cdot$sr). A similar picture can be found for other IR lines, for example [ОIII] 88~$\mu$m, [NII] 205~$\mu$m, and the dust continuum lying under them, although the excess of luminosity in these lines over the continuum is lower than that for [CII] 158~$\mu$m, and reaches only a factor of 10 -- 50 (Figure~\ref{fig-ratio-fir-lines}). The obtained values of IR intensities in the continuum and lines are quite achievable for studying SN remnants in the Galaxy and Magellanic Clouds on the high-resolution spectrograph within the planned space project “Millimetron” \citep{Kardashev2014,Novikov2021}. For the above value of the remnant brightness in the continuum, the ratio $S/N\sim 3$ is expected to be reached during the observation time of about 1~hour with the channel width of 1~MHz. Thus, the detection of IR lines can serve as a fairly good indicator of the magnitude of inhomogeneity of the medium in which the remnant is expanding. It is worth noting that with the angular resolution close to the diffraction limit for a wavelength of 158~$\mu$m $\lambda/D \sim 1.5\times 10^{-5}$, the spatial scale at distances $\simlt 10$~kpc will be less than 0.15~pc, which is better than the size of the numerical cell in the calculations. At such scales, the assumption of small spatial fluctuations in the surface brightness may not be fulfilled. The radiation from such small areas of the remnant is unlikely to reflect the properties of the remnant as a whole. Therefore, it seems more optimal to study the remnants in Magellanic Clouds.

The line width is determined by the velocity dispersion in the SN shell, or more precisely, by the expansion velocity of the SN shell \citep[see the example for optical lines in][]{vms2015,vs2024halpha}, which significantly exceeds the thermal broadening. The [CII] 158~$\mu$m and [NII] 205~$\mu$m lines arise in collisionally cooled gas with $T\simlt 2\times 10^4$~K and remain bright up to $\sim 200-300$~K, which corresponds to thermal velocities of $\sim 1-10$~km s$^{-1}$. Gas in a SN remnant has such temperature values in the dense shell, as well as inside the surviving dense fragments during the expansion of the remnant in an inhomogeneous medium with $\sigma \simgt 0.8$. In turn, the line [ОIII] 88~$\mu$m corresponds to hotter gas with $T\sim (2-20)\times 10^4$~K, which is located behind the dense shell (Figure~\ref{fig-den-maps}, left column). Thermal velocities of gas in this region of the remnant are $\sim 10-30$~km/s. The average (weighted by the gas mass) expansion velocity of the remnant changes insignificantly at SN age older than 40~kyr, but depends on the density dispersion $\sigma$ and is $\sim 60-70$~km s$^{-1}$ for $\sigma \sim 0.2$ and about $\sim 30$~km s$^{-1}$ for $\sigma \sim 2.2$. One can see thet these values are significantly higher than the thermal velocity in the dense shell and fragments, but are comparable with the values in the gas behind the shell.

\section{CONCLUSIONS}

Here we consider the evolution of the infrared luminosity of the SN remnant expanding in the inhomogeneous interstellar medium with density dispersion $\sigma$. The luminosity of the remnant is determined by the contribution from the swept-up dust emitting in the continuum and gas cooling in the lines of metal ions and atoms. The calculation of the emission properties of dust and gas is based on three-dimensional dynamics of the SN remnant in the inhomogeneous medium \citep{Dedikov2024}. The results are summarized as follows:

\begin{itemize}
 \item the mass of the swept-up interstellar dust increases rapidly during first few thousand years after the SN explosion and the total IR luminosity (in the range $\lambda = 1-1000$~$\mu$m) of the dust reaches a maximum $L_{\rm IR}\sim (4-8)\times 10^4\lsun$ and then decreases with the age of the remnant due to the destruction of particles in the hot gas and the decrease of their emissivity in the cooling gas of the shell; 
 \item with increasing inhomogeneity of the medium, the shock wave penetrates predominantly between fragments into regions of lower density, therefore the IR luminosity decreases, despite the mass of dust located in the hot gas remains almost constant before the onset of the radiative phase in most of the shell (about 40~kyr for the mean gas density $\langle n \rangle \sim 1$~cm$^{-3}$);
 \item after cooling of the gas in the remnant below $10^5$K, the decline in the IR luminosity of the dust becomes faster, e.g. for a remnant older than 60 kyr expanding in a slightly inhomogeneous medium with $\sigma \simlt 0.8$ for mean gas density of $\langle n \rangle \sim 1$~cm$^{-3}$; during evolution in more inhomogeneous medium with $\sigma \simgt 1.5$, the fraction of hot gas remains high longer, therefore the rate of decline in the IR luminosity of the remnant does not change during the period of evolution considered here;
 \item the maximum IR luminosity is achieved, when the SN remnant age is about 10~kyr, in the band with central wavelength of 40~$\mu$m and after 20 -- 30~kyr it shifts to a longer wavelength range with 70~$\mu$m, where it is located next 40 -- 50~kyr, apparently this band is optimal for studying late remnants;
 \item on the color diagrams {$I_{24}/I_{70} - I_{70}/I_{160}$} (the subscripts are the central wavelengths of the bands in $\mu$m) the spread of values becomes larger during evolution in a more inhomogeneous medium; in addition, with increasing age of the remnants, the luminosities in the shortwave bands fall more sharply and the range of values shifts noticeably on the diagram;
 \item the dust temperature changes from almost 70 to about $\sim 20$~K during the evolution and its value depends weakly on the magnitude of inhomogeneity of the medium, however, for remnants older than 40~kyr the dust temperature decreases rapidly with an increase in the average density of the medium;
 \item after the onset of radiative cooling in the massive shell of the SN remnant, strong lines of metal ions appear against the dust continuum, in particular [OIII] 88~$\mu$m, [CII] 158~$\mu$m, [NII] 205~$\mu$m and [CI] 369~$\mu$m; their luminosity grows rapidly and after saturation exceeds the luminosity of the dust in the continuum under the line by approximately $\sim 10-10^3$~times; the moment of reaching the maximum value of the luminosity in these lines depends significantly on the inhomogeneity of the medium; their relationships with each other and with respect to the dust emission make it possible to estimate the magnitude of inhomogeneity of the medium in which the remnant is expanding.

\end{itemize}

Finally, it is worth noting once again that the study of IR emission from SN remnants with an age older than 10 kyr is important for estimating the efficiency of destruction of interstellar dust by SN shocks and, consequently, for understanding the overall dust budget in galaxies \citep[e.g.,][]{McKee1989,Mattsson2011,Mattsson2021}.

\begin{acknowledgements}
The authors express their gratitude to Yu.A. Shchekinov for valuable comments and discussions, to D.I. Novikov, S.V. Pilipenko, and I.V. Tretyakov for discussions, and to I.S. Khrykin for assistance.
\end{acknowledgements}

\section*{FUNDING}

Numerical modeling of shell emission was carried out with the support of the Russian Science Foundation (grant No. 23-22-00266).

\section*{CONFLICT OF INTEREST}

The authors of this work declare that they have no conflicts interest.

\bibliographystyle{aspb1}
\bibliography{p-bib1.bib}

\end{document}